\documentclass[12pt,english]{article}
\usepackage[T1]{fontenc}
\usepackage[latin1]{inputenc}
\usepackage{a4wide}
\usepackage{amsmath}
\usepackage{babel}
\usepackage{graphics}
\usepackage{rotating}
\usepackage{amssymb}
\makeatletter

\providecommand{\LyX}{L\kern-.1667em\lower.25em\hbox{Y}\kern-.125emX\@}
\let\SF@@footnote\footnote
\def\footnote{\ifx\protect\@typeset@protect
    \expandafter\SF@@footnote
  \else
    \expandafter\SF@gobble@opt
  \fi
}
\expandafter\def\csname SF@gobble@opt \endcsname{\@ifnextchar[%]
  \SF@gobble@twobracket
  \@gobble
}
\edef\SF@gobble@opt{\noexpand\protect
  \expandafter\noexpand\csname SF@gobble@opt \endcsname}
\def\SF@gobble@twobracket[#1]#2{}

\makeatother
\begin{document}

\title{On the possiblity of detecting Solar pp-neutrino with a large volume
liquid organic scintillator detector.\thanks{Contributed paper to the Nonaccelerating New Neutrino Physics conference,
NANP-2003, Dubna. To appear in Phys.At.Nucl.(2004)}}

\maketitle

\begin{center}
\author{
\textbf{A.V.Derbin}\footnotemark[1]
\textbf{,O.Yu.Smirnov}\footnotemark[2],\footnotemark[3]
\textbf{, and O.A.Zaimidoroga}\footnotemark[3]
}
\end{center}

\footnotetext[1]{St. Petersburg Nuclear Physics Inst.itute - Gatchina, Russia.}
\footnotetext[2]{Corresponding author: E-mail: osmirnov@jinr.ru;smirnov@lngs.infn.it}
\footnotetext[3]{Joint Institutr for Nuclear Research, 141980 Dubna, Russia.}

\begin{abstract}
It is shown that a large volume liquid organic scintillator detector
with an energy resolution of 10 keV at 200 keV (\( 1\sigma  \)) will
be sensitive to solar \textit{pp}--neutrino, if operated at the target
radiopurity levels for the Borexino detector, or the solar neutrino
project of KamLAND.
\end{abstract}

\section{Introduction}

Present information on the solar neutrino spectrum is based on the
very tail of the total neutrino flux (about 0.2\%). The low energy
part of the spectrum, and in particular \textit{pp}-neutrino flux,
has not been measured directly yet. After the observation of reactor
neutrino oscillations by the KamLAND collaboration \cite{KamLAND},
the spectrometry of the low energy solar neutrinos is important for
the confirmation of the LMA MSW scenario, for the restricting of the
allowed LMA parameters region \cite{AfterKamLand}, as well as for
the search of the neutrino nonstandard properties.

The \textit{pp} neutrinos measurement is a critical test of stellar
evolution theory and of neutrino oscillation solutions. The \textit{pp}--neutrino
flux is predicted by the Standard Solar Model (SSM) with a precision
of the order of 1\%, in contrast to the 20\% precision predictions
of the high energy neutrino flux from the \( ^{8}B \). A discussion
of the physics potential of the \textit{pp} solar neutrino flux can
be found in \cite{pp_potential},\cite{WhyPP} and \cite{TheSun}.
A number of projects aiming to build \textit{pp}-neutrino spectrometers
are in the different stages of research and development.

\begin{table*}[!htb]
{\centering \begin{sideways}
\begin{tabular}{|c|c|c|c|c|c|c|}
\hline 
Project&
Method&
Threshold,&
Resolution &
Mass, t&
\multicolumn{1}{c|}{Reaction}&
SSM pp\\
(Reference)&
&
keV&
&
&
&
events,\\
&
&
&
&
&
&
\( d^{-1} \)\\
\hline 
LENS (Yb)&
\( ^{176}Yb \), &
301 (\( \nu  \))&
7\%&
20&
\multicolumn{1}{l|}{\( ^{176}Yb+\nu _{e}\rightarrow  \) }&
0.5\\
\cite{LENSYb}&
LS&
&
@ 1 MeV&
(8\% in nat \( Yb \))&
\multicolumn{1}{r|}{\( ^{176}Lu+e^{-} \)}&
\\
\hline 
LENS (In)&
\( ^{115}In \)&
118(\( \nu  \))&
18\%&
20&
\multicolumn{1}{l|}{\( ^{115}In+\nu _{e}\rightarrow  \)}&
1.4\\
\cite{LENSIn}&
LS&
&
@100 keV&
&
\multicolumn{1}{r|}{\( ^{115}Sn^{*}(613)+e^{-} \)}&
\( \epsilon (pp)=0.25 \)\\
\hline 
GENIUS&
\( ^{76}Ge \) &
11(\( e^{-} \))&
0.3\% &
1 &
\multicolumn{1}{l|}{\( \nu +e^{-}\rightarrow  \)}&
1.8\\
\cite{Genius}&
Scatt&
59(\( \nu  \))&
@ 300 keV&
10&
\multicolumn{1}{r|}{\( \nu +e^{-} \)}&
18\\
\hline 
HERON&
Superfluid \( ^{4}He \) &
50(\( e^{-} \))&
10\%&
20 (28)&
\multicolumn{1}{l|}{\( \nu +e^{-}\rightarrow  \)}&
5.5 (LMA)\\
\cite{HERON}&
Rotons/phonons+UV &
141(\( \nu  \))&
@50 keV&
&
\multicolumn{1}{r|}{\( \nu +e^{-} \)}&
\\
\hline 
XMASS&
Liquid Xe&
50(\( e^{-} \))&
11\%&
10&
\multicolumn{1}{l|}{\( \nu +e^{-}\rightarrow  \)}&
10\\
\cite{XMASS}&
Scintill&
141(\( \nu  \))&
@ 300 keV&
&
\multicolumn{1}{r|}{\( \nu +e^{-} \)}&
\\
\hline 
HELLAZ&
He (5 atm), &
100(\( e^{-} \))&
6\%&
2000 \( m^{3} \)&
\multicolumn{1}{l|}{\( \nu +e^{-}\rightarrow  \)}&
7\\
\cite{HELLAZ}&
TPC&
217(\( \nu  \))&
@800 keV&
&
\multicolumn{1}{r|}{\( \nu +e^{-} \)}&
\\
\hline 
MOON&
Drift &
168(\( \nu  \))&
12.4\% FWHH&
3.3&
\( \nu _{e}+^{100}Mo\rightarrow  \)&
1.1\\
\cite{MOON}&
Chambers&
&
@ 1 MeV&
&
\( ^{100}Tc+e^{-} \)&
\\
\hline 
MUNU&
TPC,\( CF_{4} \)&
100(\( e^{-} \))&
16\% FWHH&
0.74 &
\multicolumn{1}{l|}{\( \nu +e^{-}\rightarrow  \)}&
0.5\\
\cite{MUNU}&
Direction&
217(\( \nu  \))&
@ 1 MeV&
(200 \( m^{3} \))&
\multicolumn{1}{r|}{\( \nu +e^{-} \)}&
\\
\hline 
NEON&
He,Ne&
20(\( e^{-} \))&
16\% FWHH&
10&
\multicolumn{1}{l|}{\( \nu +e^{-}\rightarrow  \)}&
18\\
\cite{NEON}&
Scintill&
82(\( \nu  \))&
@ 100 keV&
&
\multicolumn{1}{r|}{\( \nu +e^{-} \)}&
\\
\hline 
Present&
LS&
170(\( e^{-} \))&
10.5 keV &
10&
\multicolumn{1}{l|}{\( \nu +e^{-}\rightarrow  \)}&
1.8\\
work&
&
310(\( \nu  \))&
@ 200 keV&
&
\multicolumn{1}{r|}{\( \nu +e^{-} \)}&
1.1 (LMA)\\
\hline
\end{tabular}
\end{sideways}\par}

\caption{\label{Tab:Experiments}Key characteristics of the solar neutrino
projects sensitive to the \textit{pp}--neutrino}
\end{table*}

\clearpage

The principal
characteristics of the existing proposals \cite{LENSYb}-\cite{NEON}
are shown in Table\( \:  \)\ref{Tab:Experiments}. The operating
gallium radiochemical experiments sensitive to solar \textit{pp}-
neutrinos (SAGE \cite{SAGE} and GALLEX \cite{Gallex}) are not cited
in the table, because they do not provide spectrographic information.

\begin{table*}[!htb]

\caption{\label{Table:AchivedPurity}Achived and targeted purities in the
Borexino and KamLAND solar neutrino projects.}

\( \;  \)

\( \:  \)

\begin{tabular}{|c|c|c|c|c|}
\hline 
&
CTF of Borexino&
Borexino&
KamLAND&
KamLAND \\
&
\cite{LowBckgBorexino,CTF_paper,UltraLowBkg}&
 goals, \cite{SciTechBorexino}&
\cite{KamLANDsolar}&
goals, \cite{KamLANDsolar} \\
\hline 
\( ^{14}C \)&
\( 2\times 10^{-18} \) g/g&
\( \sim 10^{-18} \) g/g&
No data&
\\
\hline 
\( ^{238}U \)&
\( <4.8\times 10^{-16} \) g/g&
\( \sim 10^{-16} \) g/g&
\( 3.5\times 10^{-18} \) g/g&
\( \sim 10^{-16} \) g/g\\
&
&
(1 \( \mu  \)Bq/m\( ^{3} \))&
&
\\
\hline 
\( ^{232}Th \)&
\( <8.4\times 10^{-16} \) g/g&
\( \sim 10^{-16} \) g/g &
\( 5.2\times 10^{-17} \) g/g&
\( \sim 10^{-16} \) g/g\\
\hline 
\( ^{40}K \)&
\( \leq 10^{-15} \) g/g &
\( \sim 10^{-18} \) g/g&
\( 2.7\times 10^{-16} \) g/g&
\( \leq 10^{-18} \) g/g\\
\hline 
\( ^{210}Pb \)&
\( <500 \) \( \mu  \)Bq/t&
\( \sim 1 \)\( \mu  \)Bq/m\( ^{3} \)&
\( \simeq 10^{-20} \) g/g &
\( 5\times 10^{-25} \) g/g\\
&
&
&
&
(1 \( \mu  \)Bq/m\( ^{3} \))\\
\hline 
\( ^{85}Kr \)&
\( <600 \) \( \mu  \)Bq/t&
\( \sim 1 \)\( \mu  \)Bq/m\( ^{3} \)&
\( 0.7 \) Bq/m\( ^{3} \)&
\( 1 \) \( \mu Bq \)/m\( ^{3} \)\\
\hline 
\( ^{39}Ar \)&
\( <800 \) \( \mu  \)Bq/t&
\( \sim 1 \)\( \mu  \)Bq/m\( ^{3} \)&
&
\\
\hline 
\( ^{222}Rn \)&
\( (3.5\pm 1.4)\times 10^{-16} \) g/g &
\( \sim 10^{-16} \) g/g&
\( 0.03 \) \( \mu  \)Bq/m\( ^{3} \)&
\( 1 \) \( \mu  \)Bq/m\( ^{3} \)\\
&
(\( \sim 3\mu  \)Bq/m\( ^{3} \))&
&
&
\\
\hline
\end{tabular}
\end{table*}

The main problem in the neutrino detection is the very small cross
sections of the neutrino interactions with matter, this demands a
large detectors with a very low intrinsic background. Below we summarize
briefly the achievements in the purification of the liquid organic
scintillators, and on the basis of the developed techniques propose
a high resolution detector, filled with a liquid organic scintillator,
with an energy threshold as low as 170-180 keV, capable of registering
solar \textit{pp}--neutrinos. A preliminary description has been reported
in \cite{SZ03}.

\section{Purities achieved with liquid scintillator detectors}

For the present moment the record on liquid organic scintillator purity
with a large scale sample has been achieved with the Borexino \cite{BorexinoProposal,SciTechBorexino}
Counting Test Facility (CTF) and KamLAND detector \cite{KamLAND}.
The available data are summarized in table \ref{Table:AchivedPurity}.
While the CTF is a prototype detector operating with 4 tones of liquid
scintillator \cite{CTF_paper,UltraLowBkg}, the KamLAND detector is
loaded with 1000 tones of the liquid scintillator. Both detectors
demonstrate very good purification of the scintillator for \( U-Th \)
and \( ^{222}Rn \). The values cited in table \ref{Table:AchivedPurity}
for the \( ^{238}U \) and \( ^{232}Th \) content are obtained by
counting the number of the decay sequences from \( ^{214}Bi \) and
\( ^{212}Bi \) in the assumption of secular equilibrium. A precise
measurement of the abundance of \( ^{40}K \) was not possible with
CTF because of the sensitivity level, but is expected to be much better
because of the high efficiency of water extraction for the removal
of \( K \) ions \cite{SciTechBorexino}. The investigation performed
in the frame of the Borexino programme shows that the content of the
\( ^{85}Kr \) and \( ^{39}Ar \) can be significantly decreased by
the proper choice of the \( N_{2} \) for the stripping. The goals
for the purity in both Borexino and KamLAND project for the observation
of the solar neutrinos are similar.

The importance of the purification of the liquid scintillator from
the \( ^{39}Ar \) and \( ^{85}Kr \) was understood during the operation
of the CTF.

\section{The design of the detector}

Even in the case that the desired purity can be achieved in the future
Borexino and KamLAND experiments, the direct measurements in the \textit{pp}--neutrino
energy region are impossible with these big detectors. In fact, the
presence of the beta-decaying \( ^{14}C \) isotope in the liquid
organic scintillator sets a lower threshold on the detector sensitivity.
The measured content of the \( ^{14}C \) in the liquid scintillator
used in the CTF detector was at the level of \( 2\times 10^{-18} \)
g/g with respect to the \( ^{12}C \) content \cite{C14_paper}, and
this is the only measurement available at such a low concentration.
Though the end point of the \( ^{14}C \) \( \beta  \)- decay is
only 156 keV, the energy resolutions of the CTF, as well as Borexino
and KamLAND, are not good enough at this energy in order to set a
threshold lower than 250 keV. 

Thus, the efforts should be concentrated on the construction of a
compact detector with the highest possible energy and spatial resolution.
We suggest to use PMTs supplied with hexagonal light concentrators
in order to provide \( 4\pi  \) coverage, in comparison to 21\% for
CTF and 30\% for Borexino and KamLAND. Additional energy resolution
improvement (about 15\%) in the low energy region can be achieved
by using an energy reconstruction technique discussed in \cite{Energy reconstruction}.
Good spatial resolution is needed in order to provide an active shielding
from the external background, mainly gammas with energy 1.45 MeV coming
from the \( ^{40}K \) decay in the PMT material. The additional passive
shielding with 200 cm of ultrapure water is considered in the present
design.

\begin{figure*}[!htb]
{\centering \resizebox*{1\textwidth}{0.5\textheight}{\includegraphics{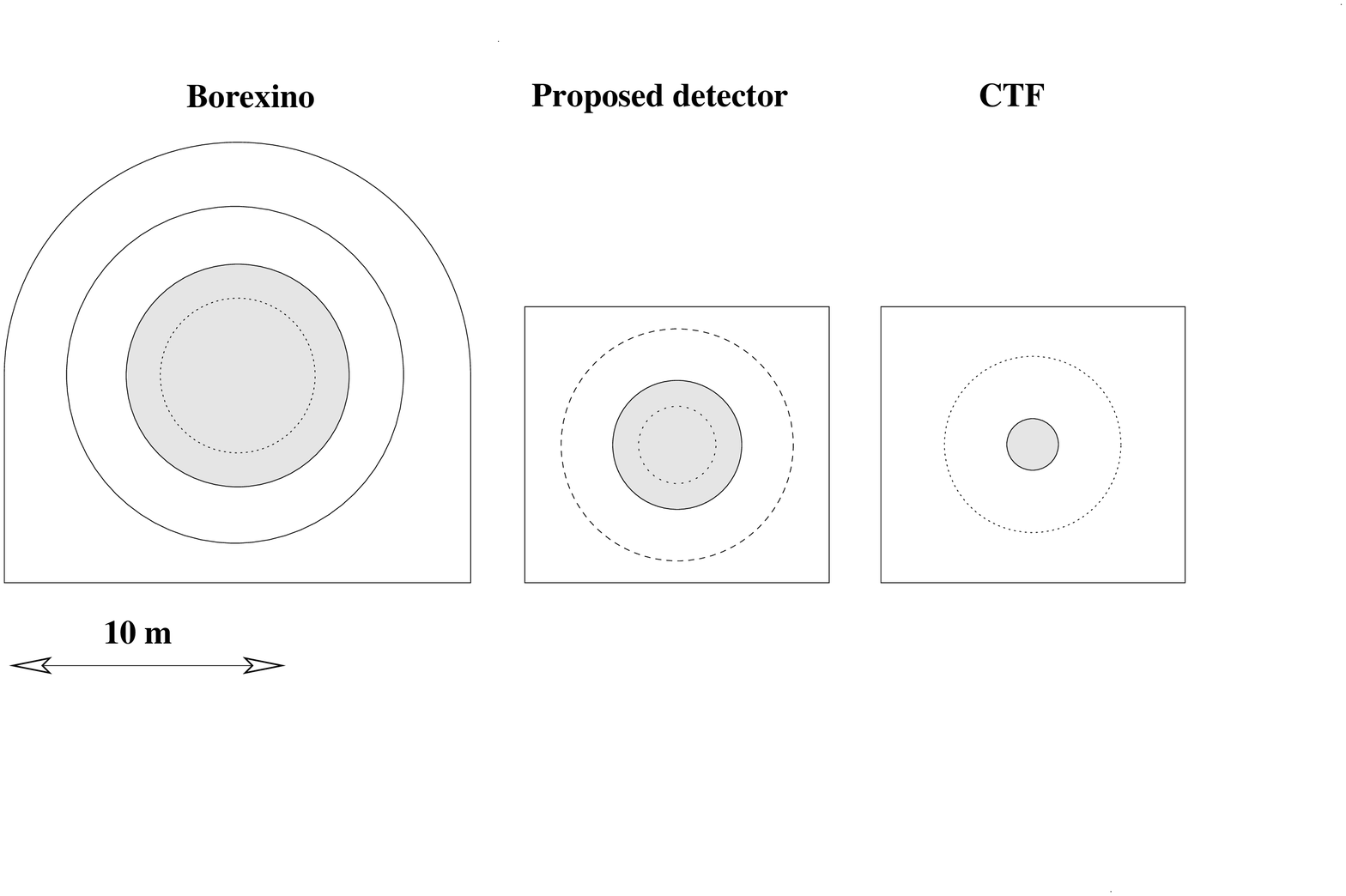}} \par}

\caption{\label{Fig:Geometry}Comparison of the geometry of the Borexino,
CTF and proposed detectors. The inner vessel with scintillator is
shown with a gray colour. The dashed line inside the inner vessel
defines the fiducial volume, the outer layer protects the fiducial
volume from the external gammas. PMTs are uniformly distributed over
the surface shown with a solid line on the Borexino drawing and with
a dashed line on the two others.}
\end{figure*}

The possible geometry of the detector is presented in the Fig.\ref{Fig:Geometry}
in comparison with Borexino and CTF sketches. The inner vessel is
a transparent spherical nylon bag with a radius of 240 cm, containing
60 tons of ultrapure pseudocumene with 1.5 g/g of PPO. The active
shielding is provided by 100 cm of the outer layer of scintillator.
The 800 PMTs are mounted on an open structure at a distance of 440
cm from the detector's center (distance is counted from the PMT photocathode).
We considered also the use of the 8'' ETL9351 series photomultiplier
\cite{PMTs}. The comparison of the geometrical parameters of Borexino,
CTF and the proposed detector is presented in the Table \ref{Table:CTF-Borex-Upgrade}.

\begin{table*}[!htb]

\caption{\label{Table:CTF-Borex-Upgrade}Comparison of the main features of
the CTF, Borexino and the proposed detector. Some data for the KamLand
detector are shown for comparison. Because of the higher threshold
(400 keV), the energy and spatial resolutions are not estimated for
KamLand.}

\( \;  \)

\( \:  \)

\( \;  \)

{\centering \begin{tabular}{|c|c|c|c|c|}
\hline 
Parameter&
CTF&
Borexino&
KamLand&
Proposed \\
&
&
&
(Solar \( \nu  \) project)&
detector\\
\hline 
Geometrical coverage&
21\%&
30\%&
34\%&
\( \cong  \)100\%\\
\hline 
Light yield (p.e./MeV)&
360&
400&
320&
1800\\
\hline 
Light yield per PMT for the event&
3.6&
0.25&
\( \ge  \)0.25&
2.25\\
 at the detector's center \( \mu _{0} \), p.e./MeV&
&
&
&
\\
\hline 
Energy resolution @ 200 keV, keV&
27&
26&
&
10.5\\
(\( \sim \frac{1}{\sqrt{Light}yield} \))&
&
&
&
\\
\hline 
Threshold, keV&
250&
250&
400&
170\\
\hline 
Muon veto PMTs&
16&
200&
&
50\\
\hline 
PMTs number&
100&
2200&
1325 (17{}``)+&
800\\
&
&
&
554 (20{}``)&
\\
\hline 
Total natural K content &
8&
176&
&
64\\
in the PMTs, g&
&
&
&
\\
\hline 
Distance between the PMTs &
330&
675&
825&
440\\
and detector's center, cm&
&
&
&
\\
\hline 
Spatial resolution @ 200 keV, cm&
20&
45&
&
8\\
(\( \sim <\frac{1}{\sqrt{N_{hit}}}>\cong \frac{1}{\sqrt{N_{PM}(1-e^{-0.2\mu _{0}})}} \))&
&
&
&
\\
\hline
\end{tabular}\par}
\end{table*}

The choice of the geometry is motivated by the following reasons:

\begin{itemize}
\item the highest possible energy and spatial resolution;
\item the detector should fit in the existing CTF external tank, which is
10 m height and 11 m in diameter;
\item the light registration system should provide maximum possible geometrical
coverage with a minimal number of PMTs required;
\item the active shielding of the fiducial volume is provided by at least
100 cm of scintillator;
\item the passive shielding against the gammas originating from the PMT
impurities is provided by 200 cm of ultrapure water;
\item the fiducial volume of the detector should be of the order of 10 tons;
\item the inner vessel size should be as small as possible in order to avoid
the loss of light in the scintillator and to provide better detector
uniformity;
\item to lower the detector's threshold (\( <35 \) keV) in order to acquire
the \( ^{14}C \) spectrum shape without deformations caused by the
threshold effects.
\end{itemize}
A larger size detector is an unfavourable solution because of the
huge number of the PMTs necessary to provide \( 4\pi  \) coverage.
The big inner vessel volume in turn decreases the amount of the light
escaping from the interior part of the detector. The spatial reconstruction
of the lower energy events is complicated because of multiple absorption
and reemission of the light on the way to the PMTs with a characteristic
length of 1 m \cite{BigDetLight}. 

The detector should be supplied with an external muon veto system.
The muon veto system consisting of about 50 additional PMTs can be
mounted on the top and on the bottom of the cylindrical external tank.
The muon recognition efficiency should be at the level of 99.99\%
in order to guarantee missed muons count \( <0.1 \) per day. The
muons flux at the LNGS underground laboratory is about 7 times less
than at the Kamioka site.

\section{Detector energy resolution}

A detailed analysis of the large volume liquid scintillator detector
energy resolution can be found in \cite{Energy reconstruction} and
\cite{Resolutions}. We give here a brief overview of the main results
because of the importance for further discussion. Taking into account
the dependence of the registered charge on energy one can write for
the CTF charge resolution%
\footnote{This is the case when no energy reconstruction is performed and the
energy is defined by dividing the total registered charge by the p.e.
yield A: \( E=\frac{Q}{A} \).
}:

\begin{equation}
\label{RES}
\frac{\sigma _{Q}}{Q}=\sqrt{\frac{1+\overline{v_{1}}}{A\cdot E\cdot f(k_{B},E)\cdot v_{f}}+v(p)},
\end{equation}

where 

\textbf{\( \overline{v_{1}}=\frac{1}{N_{PM}}\sum _{i=1}^{N_{PM}}s_{i}v_{1_{i}} \)}
is the relative variance of the PMT single photoelectron charge spectrum
(\( v_{1_{i}} \)) averaged over all CTF PMTs (\( N_{PM} \)), taking
into account the relative sensitivity \( s_{i} \) of the i-th PMT;

\textbf{\( v(p) \)} is the parameter which takes into account the
variance of the signal for the source uniformly distributed over the
detector's volume. Because of the detector's spherical symmetry one
can describe the dependence of the registered charge on the distance
from the event to the detector's center \textbf{r} with a function
\( Q(r) \) of a single parameter \textbf{r}, \mbox{\( Q(r)=Q_{0}f_{R}(r) \)},
where \( Q_{0} \) is the charge collected for an event of the same
energy occurring at the detector's center. The \( v(p) \) parameter
is the relative variance of the factor \( f_{R}(r \)) describing
the radial dependence of the registered charge:

\textbf{\begin{equation}
\label{Formula:v(p)}
v(p)\equiv \frac{<f_{R}^{2}(r)>_{V}}{<f_{R}(r)>^{2}_{V}}-1;
\end{equation}
}

\textbf{\( v_{f} \)} is volume factor, coming from the averaging
of the signals over the CTF volume, \( v_{f}\equiv \frac{<Q(r)>_{V}}{Q_{0}} \).

\textbf{\( f(k_{B},E) \)} is a function taking into account the suppression
of the light yield at low energies, the so called ionization quenching.

The coefficient \textbf{\( A \)} linking the event energy and the
total collected charge is called the light (or photoelectron) yield.
The light yield for electrons can be considered linear with respect
to its energy only for energies above 1 MeV. At low energies the phenomenon
of ``ionization quenching'' violates the linear dependence of the
light yield versus energy \cite{Birks}. The deviations from the linear
law can be taken into account by the ionization deficit function \( f(k_{B},E) \),
where \( k_{B} \) is Birks' constant. For the calculations of the
ionization quenching effect for PC scintillator we used the KB program
from the CPC library \cite{Quenching}.

For the details of the meaning of the parameters see \cite{Energy reconstruction}
and \cite{Resolutions}. For the signal calculation we used the following
parameters: \( A=1800 \)\( \:  \)p.e./MeV, \( v_{1}=0 \), \( v_{f}=1 \),
\( v(p)=2.3\times 10^{-3} \), \( k_{B}=0.0167 \). 

The signal \( S(Q) \) registered by the detector is the convolution
of the {}``pure'' signal spectrum \( S_{0}(Q) \) with the detector's
resolution: 

\begin{equation}
\label{Model}
S(Q)=N_{0}\int S_{0}(E(Q'))\frac{dE}{dQ}Res(Q,Q')dQ'
\end{equation}

where \( Res(Q,Q')=\frac{1}{\sqrt{2\pi }\sigma _{Q}}e^{-\frac{1}{2}(\frac{Q-Q'}{\sigma _{Q}})^{2}} \)
is the detector response function, and \( \sigma _{Q} \) is defined
by (\ref{RES}).

\section{Backgrounds}

The sensitivity of the detector to the solar \textit{pp}-neutrino
depends on the background level in the 170-- 250- keV energy window.
As in Borexino, CTF (\cite{BorexinoProposal}, \cite{CTF_paper})
and the KamLAND solar neutrino project \cite{KamLANDsolar}, the main
sources of background are

\begin{itemize}
\item internal background, including \( ^{14}C \) beta- decay counts in
the neutrino window;
\item background from the radon dissolved in the buffer ;
\item external gamma background;
\item cosmic ray background.
\end{itemize}
We considered the contamination of the liquid scintillator with the
radionuclides on the levels given in table \ref{Table:AchivedPurity}
for Borexino. 

In the following subsections we give an estimate of the background
contribution from each source, showing that the main contribution
into the background is due to the internal radionuclide decays. The
Monte- Carlo method has been used in our calculations in order to
simulate the detector response to the background events. The code
is split in two parts: the electron-gamma shower simulation (EG code)
and the simulation of the registered charge and position (REG code).
The EG code generates a random position event with a random initial
direction (for gammas) and follows the gamma- electron shower using
the EGS-4 code\cite{EGS4}. The electrons and alphas are not propagated
in the program and are considered to be point-like, with the position
at the initial coordinates. The mean registered charge corresponding
to the electron's energy \( E_{e} \) is calculated by

\begin{equation}
\label{Qe}
Q_{e}=A\cdot E_{e}\cdot f(k_{B},E)\cdot f_{R}(r),
\end{equation}

where \( f_{R}(r) \) is a factor, taking into account the dependence
of the registered charge on the distance from the detector's center,
and \( f(k_{B},E_{e}) \) is the quenching factor for electrons. 

The factor \( f_{R}(r) \) was estimated with the Monte Carlo method,
simulating the light collection from the source, placed at different
distances from the detector's center. The quenching factor \( k_{B}=0.0167 \)
was independently measured for the scintillator on the base of pseudocumene
(PC) \cite{Peron}. The value is in agreement with a high statistics
fit of the \( ^{14}C \) \( \beta - \)spectrum of the CTF data. 

The mean registered charge corresponding to an alpha of energy \( E \)
is calculated by

\begin{equation}
\label{Qalpha}
Q_{\alpha }=A\cdot E_{\alpha }\cdot f_{\alpha }(E)\cdot f_{R}(r),
\end{equation}

where \( f_{\alpha }(E) \) is the quenching factor for alphas. The
following approximation of the quenching factor\( f_{\alpha }(E) \)
was used for the simulations with PC \cite{SBonetti}:

\[
f_{\alpha }(E)=\frac{1}{20.3-1.3\cdot E},\]
where alpha energy \( E \) is measured in MeV.

The gammas were propagated using the EGS-4 code. As soon as an electron
of energy \( E_{e} \) appears inside the scintillator, the corresponding
charge is calculated:

\begin{equation}
\label{DeltaQ}
\Delta Q_{i}=A\cdot E_{e_{i}}\cdot f(k_{B},E_{e})\cdot f_{R}(r_{i}).
\end{equation}

The total mean collected charge is defined when the gamma is discarded
by the EG code, summing individual deposits:

\[
Q_{\gamma }=\sum \Delta Q_{i}.\]

The weighted position is assigned to the final gamma:

\begin{equation}
\label{Formula:xw}
x_{w}=\frac{\sum \Delta Q_{i}\cdot x_{i}}{\sum \Delta Q_{i}},
\end{equation}

where \( \Delta Q_{i} \) and \( x_{i} \) are the charge deposited
for the \( i- \)th electron at the position \( \{x_{i},y_{i},z_{i}\} \).
The same rule is applied for the \( y_{w} \) and \( z_{w} \) coordinates. 

In the next step a random charge is generated according to the normal
distribution with a mean value of \( Q=\sum \Delta Q \) and with
variance \( \sigma _{Q}=\sqrt{(1+\overline{v_{1}})\cdot Q} \). Finally,
the radial reconstruction is simulated taking into account the energy
dependence of the reconstruction precision. It is assumed that the
reconstruction precision is defined by the number of PMTs fired in
an event, and that the reconstruction precision doesn't depend on
the position. These two facts were confirmed by the measurements with
an artificial radon source inserted in the CTF-I and CTF-II detectors
\cite{UltraLowBkg}. The reconstruction precision for the radon events
can be obtained either by the direct measurement with a source or
by fitting the distribution of the radon events. The mean number of
channels fired for an event with an energy E at the detector's center
is:

\begin{equation}
\label{Formula:<N>}
<N>=N_{PMT}(1-e^{-\mu _{0}}),
\end{equation}

where \( \mu _{0} \) is the mean number of photoelectrons registered
by one PMT in the event and \( N_{PMT} \) is the total number of
the PMTs. If we assume that the reconstruction precision is defined
by the mean number of fired channels, then the reconstruction precision
is:

\begin{equation}
\label{Formula:RecPrecision}
\sigma _{R}(E)=\sigma _{R}(E_{ref})\cdot \sqrt{\frac{<N(E_{ref})>}{<N(E)>}},
\end{equation}

where \( \sigma _{R}(E_{ref}) \) is the spatial resolution for a
monoenergetic source with energy \( E_{ref} \). We used as reference
the values obtained with the CTF-I detector with a \( ^{214}Po \)
source: \( \sigma _{R}(0.751\: keV)=12.3\pm 0.04 \) cm \cite{UltraLowBkg}.

We expect less then 1 events per day due to the internal background
in 10 tons of scintillator in the energy window 170-250 keV. The better
energy and spatial resolutions of the detector will permit us to improve
the \( \alpha /\beta  \) discrimination capability in comparison
to CTF. The very low energy threshold together with better energy
and spatial resolutions will allow as well to improve the selection
of the sequential decays from the radioactive chains.

\subsection{Internal background from the metallic ions and radioactive noble
gases}

The contamination of the scintillator with natural radioactive isotopes
gives a total rate of 1320 events/year in the energy window 170-750
keV with the following assumptions:

\begin{itemize}
\item the content of radioactive isotopes in the scintillator is given by
table \ref{Table:AchivedPurity}; 
\item secular equilibrium of the radioactive elements in the decay chains;
\item a 95\% capability to reject alphas (\( \alpha /\beta  \)- discrimination
technique based on the different shape of the detector response to
\( \alpha  \) and \( \beta  \));
\item 95\% rejection efficiency of the delayed coincidence method based
on the tagging of the \( ^{214}Bi-^{214}Po \) decay chain;
\item 95\% efficiency of the statistical subtraction method based on the
deducing the isotopes in the \( Rn \) chain preceding the Bi--Po
coincidence.
\end{itemize}
A more complete discussion of the background reduction techniques
can be found in \cite{BorexinoProposal}. The excellent detector's
resolution can improve the efficiency of all the techniques. 

Only 230 events of the total amount falls into the 170-- 250- keV
energy window.

\subsection{Internal background from \protect\( ^{14}C\protect \) decays}

\subsubsection{\protect\( ^{14}C\protect \) spectrum}

The major part of the background in liquid organic scintillators in
the energy region up to 200 keV is the \( \beta  \)-activity of \( ^{14}C \).
The \( \beta  \)-decay of \( ^{14}C \) is an allowed ground-state
to ground-state (\( 0^{+}\rightarrow 1^{+} \)) Gamow-- Teller transition
with an endpoint energy of \( E_{0}=156 \)\( \;  \)keV and half
life of 5730 yr. For the evaluation of the \( ^{14}C \) background
we used the \( \beta  \)- energy spectrum with a massless neutrino
in the form \cite{Morita}:

\begin{equation}
\label{BetaSpectrum}
dN(E)\sim F(Z,E)C(E)pE(Q-E)^{2}dE
\end{equation}

where 

\textbf{\( E \)} and \textbf{\( P \)} are the total electron energy
and momentum; 

\textbf{\( F(E,Z) \)} is the Fermi function with correction of screening
by atomic electrons; 

\textbf{\( C(E) \)} contains departures from allowed shape. 

For \( F(E,Z) \) we used the function from \cite{Simpson} which
agrees with tabulated values of the relativistic calculation \cite{RElativistic}.
A screening correction has been made using Rose's method \cite{Rose}
with screening potential \( V_{0}=495 \)\( \;  \)eV. The \( ^{14}C \)
spectrum shape factor can be parametrized as \( C(E)=1+\alpha E \).
In our calculation we used the value \( \alpha =-0.72 \) \cite{C14_paper}. 

The total amount of the events in the 172-- 250- keV energy window,
with the energy resolution corresponding to 1800 p.e./MeV, is 1500
ev/y/10 t if the \( ^{14}C \) content is \( 2\times 10^{-18} \)
g/g. This is the content measured at the CTF-I setup \cite{C14_paper}.

\subsubsection{\protect\( ^{14}C\protect \) spectrum and the detector's threshold}

In order to separate events from the background near the \( ^{14}C \)
spectrum end point, it is necessary to acquire the part of the spectrum
under the physical threshold of the detector (170 keV). We propose
to use the following technique for the detector triggering. First,
the lower level trigger is produced as a coincidence of the signals
from 20 PMTs in a 50 ns gate. This will give a negligible random coincidence
rate at the level \( <10^{-10} \) ev/y if all the PMTs have a dark
rate less than 5 kHz. The high level trigger is produced if the total
collected charge is greater than the preset threshold \( Q_{th} \).
The choice of this threshold will be defined by the resolution of
the detector. Let us estimate the last quantity. The mean number of
channels fired for an event with an energy E at the detector's center
can be calculated using (\ref{Formula:<N>}) with \( N_{PM}=800 \).
The solution of (\ref{Formula:<N>})
for \( <N>=20 \) will yield \( \mu _{0}\simeq 0.025 \), i.e. a total
collected charge of 20 p.e. This value is the detector threshold in
the sense that only 50\% of the events with an energy corresponding
to this charge are registered. Of course, this causes a significant
spectrum deformation near the threshold. In order to avoid these deformations
one should set the threshold at a level that will cut the events with
energies that are not providing 100\% registration, i.e. \( Q_{th}+3\sigma _{Q_{th}}=20+3.\sqrt{20}=33.4 \)\( \;  \)p.e.
This charge corresponds to approximately 35 keV if the ionization
quenching at this energy is 50\%. Though this calculation was performed
for an event at the detector's center, and the real situation is complicated
by the electronics threshold, it gives a value very close to the one
obtained with the Monte Carlo simulation.

\begin{table*}[!!!tb]

\caption{\label{Table:C14}The effect of the \protect\( ^{14}C\protect \)
in the scintillator on the sensitivity of the detector to the SSM
\textit{pp}-- neutrinos (in LMA MSW scenario). The data corresponds
to a detectors mass of 10 tones and for 1 year of the data taking.}

\( \:  \)

\( \:  \)

\( \:  \)

{\centering \begin{tabular}{|c|c|c|c|c|}
\hline 
\( ^{14}C \), g/g&
\( 2\times 10^{-18} \)&
\( 10^{-19} \)&
\( 10^{-20} \)&
\( 10^{-21} \)\\
\hline 
Threshold (\( \sqrt{bkg} \)=eff)&
172 (40)&
152&
108&
0\\
\hline 
Threshold (2\( \sqrt{bkg} \)=eff)&
178&
163&
140&
0\\
\hline 
Energy interval&
172--250&
150--250&
150-250&
150--250\\
\hline 
\( ^{14}C \) events &
1500 &
5383&
538&
54\\
\hline 
Internal Background&
228&
287&
287&
287\\
\hline 
pp (LMA)&
412&
705&
705&
705\\
\hline 
Total \( \nu  \)'s (LMA)&
668&
1035&
1035&
1035\\
\hline
\end{tabular}\par}
\end{table*}

It is commonly assumed that the \( ^{14}C \) content sets the limit
on the sensitivity at the low energy region in liquid organic scintillators.
A ratio as low as \( 2\times 10^{-18} \) g/g was achieved with CTF
detector. There are indications that the content of \( ^{14}C \)
can be even smaller, of the order of \( 10^{-21} \) g/g \cite{ElisaTesi}%
\footnote{New petro-geological models allow such a low value; contamination
with modern \( ^{14}C \) in this case have to be excluded during
petroleum refinement \cite{ElisaTesi}. The existing CTF setup is
a suitable device for the search of the organic LS with minimal \( ^{14}C \)
contamination.
}. In this case the \( ^{14}C \) contribution in the background can
be reduced by a factor 2000. It is interesting to study the dependence
on the content of the \( ^{14}C \) in the scintillator for the sensitivity
of the detector to the pp-- neutrinos. The results of the study are
summarized in Table \ref{Table:C14}. One can see that the detector
sensitivity varies rather slowly with the decrease of the \( ^{14}C \)
content in the scintillator. There are several reasons for this behaviour.
First of all, the pp-- neutrino rate is quite low and with minimal
background contribution the statistical fluctuations of the pp- rate
are the major source of the uncertainty. Another source of uncertainty
is the irreduceable internal background which becomes comparable to
the \( ^{14}C \) events contribution with the lower \( ^{14}C \)
content. The last reason is the lower threshold of the detector of
about 25 keV, which can't be decreased without increasing the random
electronics noise. In order to avoid the influence of the threshold
effect on the spectrum shape it is necessary to set the software threshold
even higher, to about 40 keV. Another motivation to set a higher software
threshold is the presence of low energy external gammas which can
be reconstructed inside the fiducial volume due to the poor spatial
resolution at low energies.

We can conclude that lower \( ^{14}C \) content would be desirable
but is not critical for the detector sensitivity to \textit{pp}--
neutrinos.

\subsubsection{\protect\( ^{14}C\protect \) pile-up events}

A potential danger are the \( ^{14}C \) pile-up events, i.e. events
occurring sequentially within a coincidence window. Such events can
in principle influence the \( ^{14}C \) spectrum tail. The fraction
of the \( ^{14}C \) pile-up events with energies above 170\( \:  \)keV
is about 5\%. The total amount of pile-up events depends on the \( ^{14}C \)
relative abundance and on the coincidence window:

\begin{equation}
\label{Formula:PileUpRate}
N_{p.u.}=\tau _{Gate}\times f_{14_{C}}^{2}\times T,
\end{equation}

where T is the total time of the data taking, \( \tau _{Gate}=60\: ns \)
is the coincidence gate width, and \( f_{14_{C}} \) is the frequency
of the \( ^{14}C \) events. For a \( ^{14}C \) abundance of \( 2\times 10^{-18} \)
g/g, the mean rate of the events caused by \( ^{14}C \) \( \beta  \)-
decay is 2.2\( \:  \)Hz. With these values the number of pile-up
events is 2.5 per day, and the number of events with energy greater
then 170 keV is only about 0.13 per day. 

The selection of point-like events provides a further possibility
to suppress the amount of these events by a factor of at least \( \left( \frac{\frac{4}{3}\pi (3\sigma _{R})^{3}}{V_{FV}}\right) ^{2}=\left( \frac{3\sigma _{R}}{R_{FV}}\right) ^{6}\simeq 10^{-5} \),
where \( \sigma _{R} \) is the spatial resolution at 170\( \:  \)\( \:  \)keV
and \( R_{FV}=140\: \: cm \) is the radius of the fiducial volume.
Thus, one can conclude that pile-up events will not influence the
shape of the \( ^{14}C \) spectrum within the considered energy interval.

\subsection{External gamma background}

The external background counts are caused mainly by the radioactive
contamination of the PMT glass with \( ^{40}K \) and elements of
the U-Th chain. The assumed content of the \( ^{238}U \), \( ^{232}Th \)
and \( K_{nat} \) in the PMTs is \( 112\; \mu  \)g/PMT, \( 47\; \mu  \)g/PMT
and \( 62\; m \)g/PMT respectively, which corresponds to the measured
radioactive contamination of the phototubes produced with high purity
glass \cite{LowBckgBorexino}. We add 30\% to these values to account
for the radioactive contamination of the concentrator and PMT divider,
sealing and support structure. Another source of external gammas is
radon dissolved in the water buffer.

The results of a simulation show that a R<150 cm spatial cut will
eliminate all the events in the neutrino window (170--250- keV). Nevertheless,
in order to reduce the background from the penetrating gamma's we
suggest to reduce the amount of the construction materials contributing
to the background. A significant amount of the material in Borexino
is contained in the mu-metal shield of the PMTs, which provides the
screening of the PMTs against the terrestrial magnetic field. An alternative
solution based on the PMTs orientation has been studied in \cite{Mu-metal}.
The effect of the PMTs orientation is comparable to the one achieved
with the PMT screening with the high magnetic permeability metal.
Use of this technique could eliminate about 1 kg of material for each
PMT in proximity to the inner vessel.

Another possibility to reduce the gamma background assumes the use
of a different topology of the events produced by the electrons and
gammas. The excellent spatial resolution of the detector will permit
distinguishing point-like energy deposits for electrons from the distributed
gamma events (\( \beta /\gamma  \) discrimination). The study of
the possibility of such discrimination for the Borexino detector is
now in progress.

\subsection{Cosmic ray induced background}

The cosmic ray induced background can be subdivided into the three
categories:

\begin{enumerate}
\item Muons crossing the water buffer of the detector, producing Cerenkov
light;
\item Neutrons produced by muon interactions, and sequentially stopped in
the water or the scintillator and emitting a 2.2 MeV annihilation
gamma;
\item Secondary radioactive nuclei produced in the muon interactions inside
the detector.
\end{enumerate}
Most of the background counts associated with muons can be effectively
removed by the muon identifying system (muon veto). One can use the
time and spatial structure of the muon induced events in order to
recognize them. The muon identification procedure was able to recognize
95\% of the muon induced events in the CTF-I detector \cite{CTF_paper}.
We assume also the use of a set of PMTs situated on the top and bottom
of the cylindrical external tank that will increase the muon identification
to a value approaching 100\%.

Some of the radioactive products of the muon interactions with the
scintillator have significant life times, that makes it impossible
to use a muon tag. These isotopes are \( ^{11}Be \) (\( \beta ^{-} \),
11.5\( \:  \)MeV, 13.8\( \:  \)s), \( ^{10}C \) (\( \beta ^{+}, \)
1.9\( \:  \)MeV + \( \gamma  \),0.72\( \:  \)MeV, 19.3 s), \( ^{11}C \)
(\( \beta ^{+} \), 0.99\( \:  \)MeV, 20.38\( \:  \)min) and \( ^{7}Be \)
(\( \gamma , \) 0.478\( \:  \)MeV, 53.3\( \:  \)d). The considered
neutrino window is too narrow to pick up a significant amount of events
from these isotopes. Precise evaluations are now in progress for the
Borexino detector \cite{Muons-induced}, but this background will
certainly be negligible in comparison to the other sources considered.

\section{Neutrino signals}

In the calculations we used SSM fluxes given by the standard solar
model\cite{BP2000}, neutrino energy spectra from \cite{CNO}--\cite{pp-hep}
and survival probabilities for the LMA solar neutrino scenario from
\cite{MSW survival}. Signal shapes were convolved with the detectors
response function using (\ref{Model}).

\subsection{Sensitivity to the \textit{pp}--neutrinos.}

The expected \textit{pp}--neutrino count for the LMA solar neutrino
oscillation scenario is listed in table \ref{Table:C14}. The sensitivity
was estimated with the Monte Carlo (MC) method. First, the total signal
was calculated taking into account the detector's resolution. In the
next step the normally distributed random signal was generated at
each bin. 

A fitting function consists of a function with 3 contributes describing
the internal background (without \( ^{14}C) \), the spectrum of the
\( ^{14}C \) decay and the neutrino signal:

\begin{equation}
\label{Formula:FittingFunc}
f(q)=N_{Bkg}Bkg(q)+N_{\, ^{14}C}\cdot C(q)+N_{\nu }\phi _{\nu }(q)
\end{equation}

The shape of the internal background Bkg(q) was fixed, but its normalization
(\( N_{Bkg} \)) was free. Another free parameter is the normalization
of the \( ^{14}C \) spectrum \( N_{\: ^{14}C} \). 

\begin{figure*}[!htb]
{\centering \resizebox*{1\textwidth}{0.5\textheight}{\includegraphics{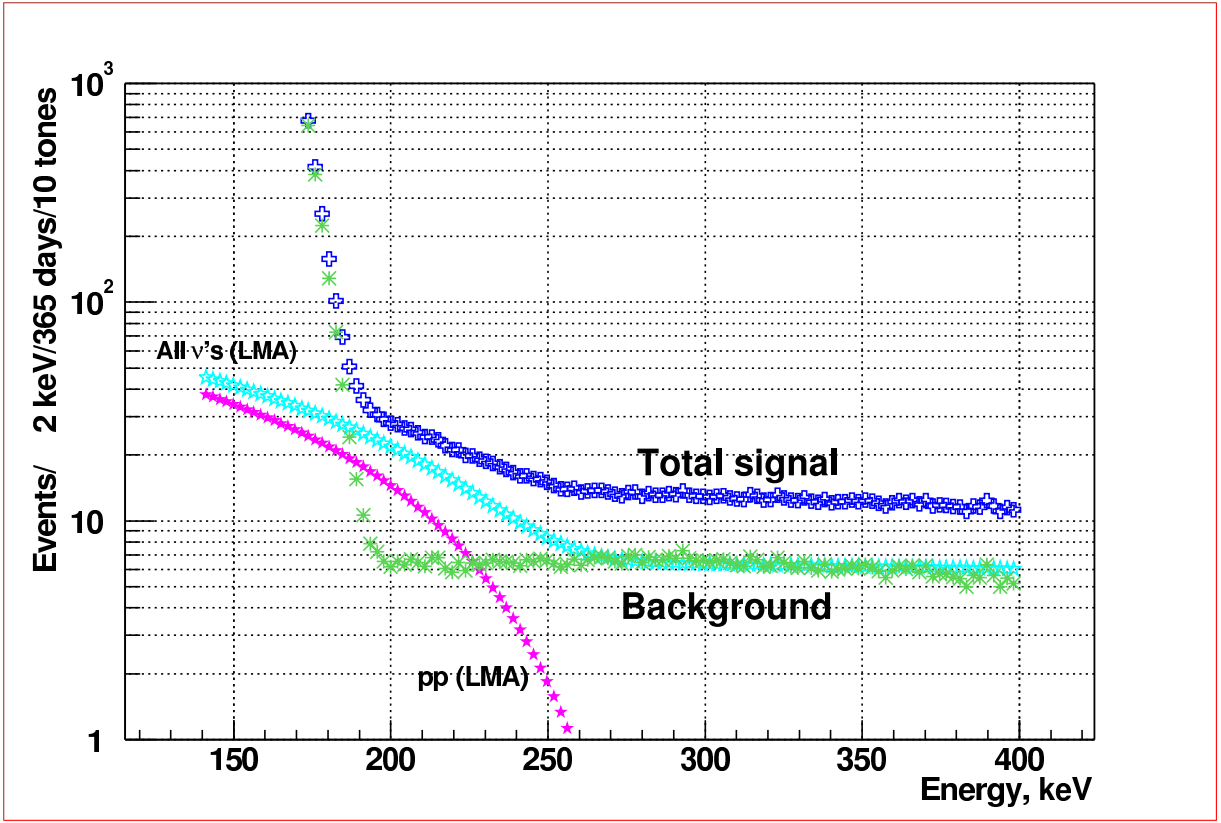}} \par}

\caption{\label{Fig:EW}Signal and background shape for the SSM neutrino fluxes
in the LMA MSW solution. The \protect\( ^{14}C\protect \) content
is \protect\( 2\times 10^{-18}g/g\protect \). The concentrations
of the main contributors to the background are listed in table. The
detectors mass is 10 tons. The resolution is calculated with the assumption
of 100\% geometrical coverage using CTF-I light output for the liquid
scintillator (i.e. 1800 p.e./MeV) and is assumed to be \protect\( \frac{1}{\sqrt{N_{p.e.}}}\protect \).
Shown signals correspond to 1 year of the data taking.}
\end{figure*}

The expected rates are listed in Table \ref{Table:C14}. The neutral
current channel for the neutrinos of non- electron flavours are taken
into account in the calculations. Other neutrino sources also have
non- negligible contributions to the total signal in this energy window.
The main source besides the \textit{pp} are \( ^{7}Be \) neutrinos
with a flat spectrum (see Fig.\ref{Fig:EW}).

Using model (\ref{Formula:FittingFunc}) we find out that the total
neutrino flux will be measured with 7.5\% (1\( \sigma  \)) relative
precision. Possible systematics errors due to the unknown shape of
the background are not included in the estimation. We assume that
the MC simulation can reproduce the form of the background and only
the total normalization of this shape is unknown. The assumption is
reasonable because of the quite narrow signal window, where the background
is dominated by the slowly varying continuous spectrum of the soft
part of the gamma spectra of radioactive impurities.

\subsection{Sensitivity to \protect\( ^{7}Be\protect \) neutrinos}

The detector will count 1070 \( ^{7}Be \) SSM LMA neutrinos per year
in the 200--700- keV energy window with an internal background of
1130 events. For comparison, the Borexino detector will count 9390
events in the 250-750 keV energy window with a background of \( \sim  \)10500
events. We are not presenting here the evaluation of the sensitivity
of the detector to the \( ^{7}Be \) neutrinos. It is clear that the
lower mass (factor 10) with comparison to the Borexino detector will
limit the sensitivity. A certain gain in the sensitivity can be achieved
due to better energy resolution of the detector (factor 2.1). The
sensitivity relative to Borexino for equal time of data taking and
equal specific background can be estimated as \( \sqrt{\frac{M_{Det}}{M_{Borex}}\frac{\sigma _{Borex}}{\sigma _{Det}}}\simeq 0.45 \)
for the measurements on the edge of the recoil electrons of the \( ^{7}Be \)
neutrino (about 660 keV).

\section{\label{Improvement}Improvement of the detector performances}

The performance of the detector can be improved by using any of the
following ideas:

\begin{enumerate}
\item \textbf{Use of the specially designed photomultipliers, providing
better quantum efficiency}. The basic idea is the {}``recycling''
of the incoming photons. Various optical arrangements have been used
to improve light absorption by letting incoming light interact with
the photocathode material more than once (see i.e. \cite{Optical}).
The idea has been revived in recent works \cite{Realisation},\cite{S20},
where the authors reported significant increase of the quantum efficiency,
up to a factor 2. There are also indications on the possibility of
creating a photocathode with very high quantum efficiency using a
material doped with nanoparticles \cite{Patent}.
\item \textbf{Use of the beta/gamma discrimination techniques}. The use
of a different topology of the point-like beta events and the spatially
distributed gamma- events can provide an opportunity to discriminate
between beta and gamma induced signals with high efficiency. The method
exploits the superior resolutions of the detector.
\item \textbf{Choice of the organic scintillator with lower content of \( ^{14}C \).}
There are indications that the content of \( ^{14}C \) can be much
smaller than measured with the CTF-I detector, namely of the order
of \( 10^{-21} \) g/g \cite{ElisaTesi}. In this case the \( ^{14}C \)
contribution in the background can be significantly reduced, and will
lead to an improvement of the detector's characteristics. 
\end{enumerate}

\section{Conclusions}

It is shown that a high energy resolution detector with the radiopurity
levels necessary for the operation of Borexino, as well as solar neutrino
project of KamLAND, will be sensitive to solar \textit{pp}--neutrinos.
The project can compete with other existing proposals (see Table\( \:  \)\ref{Tab:Experiments}). 

\( \:  \)

This job would have been impossible without the support from the INFN
sez. di Milano. We would like to thank Prof. G.Bellini and Dr. G.Ranucci
for the continuos interest in our job. 

We would like to thank all the colleagues from the Borexino collaboration
for the pleasure to work together. Special thanks to R.Ford for the
careful reading of the manuscript.

\end{document}